\documentclass[preprint, 10pt]{elsarticle}
\biboptions{sort&compress}

\usepackage[utf8]{inputenc} 
\usepackage[T1]{fontenc}    
\usepackage{xr-hyper}
\usepackage{url}            
\usepackage{booktabs}       
\usepackage{amsfonts, amsmath, amssymb, amsthm}
\usepackage{microtype}      
\usepackage{graphicx}
\usepackage{xcolor}
\usepackage{float}
\usepackage{subfiles}
\usepackage{subcaption}
\usepackage{hyperref}     
\usepackage{xspace}
\usepackage{enumitem}
\usepackage{breakurl}

\newcommand{\modelnm}{Simulation for Infection Control Operations\xspace}
\newcommand{\modelabrev}{SICO\xspace}

\theoremstyle{plain}      
\theoremstyle{plain}      \newtheorem*{thm*}{Theorem}
\theoremstyle{plain}      
\theoremstyle{plain}      
\theoremstyle{plain}      \newtheorem*{lma*}{Lemma}
\theoremstyle{plain}      
\theoremstyle{plain}      \newtheorem*{cor*}{Corollary}
\theoremstyle{plain}      
\theoremstyle{plain}      \newtheorem*{claim*}{Claim}
\theoremstyle{plain}      
\theoremstyle{plain}      \newtheorem*{prop*}{Proposition}
\theoremstyle{plain}      
\theoremstyle{plain}      
\theoremstyle{remark}     
\theoremstyle{remark}     \newtheorem*{rmk*}{Remark}
\theoremstyle{remark}     
\theoremstyle{definition} 
\theoremstyle{definition} \newtheorem*{defn*}{Definition}
\theoremstyle{definition} 
\theoremstyle{definition} \newtheorem*{ex*}{Example}
\theoremstyle{plain}

\journal{Journal of Computational Science}
\begin{document}

\externaldocument{supp}

\begin{frontmatter}

\title{\modelabrev: \modelnm }

\author[l1]{Karleigh Pine\corref{cor1}}
\ead{karleigh.pine@matrixresearch.com}
\author[l2]{Razvan Veliche\fnref{fn1}}
\ead{razvan.veliche@gmail.com}
\author[l3]{Jared Bennett}
\ead{jbennett@mobiuslogic.com}
\author[l1]{Joel Klipfel}
\ead{joel.klipfel@matrixresearch.com}

\cortext[cor1]{Corresponding author}
\fntext[fn1]{Work completed while affiliated with M\"obius Logic.}
\affiliation[l1]{organization={Matrix Research}, addressline={3844 Research Blvd.}, city={Beavercreek}, postcode={45430}, state={OH},country={USA}}
\affiliation[l2]{organization={Keystone Strategy LLC}, addressline={116 Huntington Ave.}, city={Boston}, state={MA}, postcode={02116}, country={USA}}
\affiliation[l3]{organization={M\"obius Logic}, addressline={1775 Tysons Blvd.}, city={Tysons}, state={VA}, postcode={22102}, country={USA}}

\begin{abstract}
In response to the COVID-19 pandemic and the potential threat of future 
epidemics caused by novel viruses, we developed a flexible framework 
for modeling disease intervention effects. This tool is intended to aid
decision makers at multiple levels as they compare possible responses to emerging 
epidemiological threats for optimal control and reduction of harm. The framework is 
specifically designed to be both scalable and modular, allowing it to 
model a variety of population levels, viruses, testing methods and 
strategies--including pooled testing--and intervention strategies.
In this paper, we provide an overview of this framework and examine the impact of 
different intervention strategies and their impact on infection dynamics. 
\end{abstract}

\end{frontmatter}


\section{Introduction}
\label{sec:introduction}

COVID-19 emerged from obscurity and rapidly became the most destructive event 
of the century. The death toll currently stands at nearly 7 million lives 
lost~\cite{WHO}. The economic impact exceeds 3.9\% of the median global 
GDP~\cite{econ_2022} and is expected to slow global economic recovery for the next 
several years. The societal costs of lockdowns and stay-at-home orders will be 
years manifesting. To achieve and maintain a sense of normalcy in the wake of 
COVID-19, we must find methods to surveil emerging variants and limit further 
transmission, thereby preventing additional waves of infection and potential 
lockdown situations. Moving forward from COVID-19, our goal is resiliency 
against future pandemics~\cite{marani_intensity_2021} with the development of 
flexible technologies that can be adapted to different disease and population 
characteristics. Expedient modeling of infectious disease transmission mitigation
measures can inform decision makers in the critical early days of infection spread.

The work of Lyng \textit{et al.}~\cite{lyng_2021} and Augenblick 
\textit{et al.}~\cite{augenblick_2022} has demonstrated the importance of 
modeling many combinations of testing and intervention strategies over time. 
Additionally, a dynamic approach to modeling is needed as infection rates and 
population characteristics fluctuate. In practice, the disease management 
problem is complex, with many possible population-dependent intervention 
alternatives. Therefore, it is necessary to model many aspects of infection and 
population dynamics to effectively determine an ideal disease control strategy. 
We address this need with the creation of our \modelnm{} (\modelabrev{}) that 
allows users to explore alternative planning 
and testing scenarios, adapting situational parameters such as vaccination rate, 
isolation, and testing with customization based on the disease scenario being 
considered. The tool is built utilizing an agent-based model (ABM) which 
provides flexibility in specifying the disease dynamics, test availability, and 
economic constraints important for its application in the management of future 
pandemics~\cite{marani_intensity_2021}. The use of an ABM allows \modelabrev{} 
to account for variations in agent behaviors, such as propensity to vaccinate 
and likelihood to self-isolate upon symptoms. Further, \modelabrev{} uses a 
stochastic ABM to simulate various infectious disease intervention 
strategies\textemdash{}thus enabling a decision maker to choose an optimal 
strategy for long-term disease reduction which adheres to any physical 
constraints (\textit{e.g.} test or vaccine availability) they face.   

In introducing \modelabrev{}, our contribution to the field is developing a 
hybrid epidemilogical ABM with a flexible, compartmentalized and modular design 
which allows it to be adapted and customized to fit a variety of infectious 
diseases, disease propagation scenarios and intervention strategies. Some 
noteworthy features of 
\modelabrev{} include:
\begin{itemize}
	\item Ability to model various vaccination strategies based on 
		vaccine supply and individual agents' propensity to vaccinate;
	\item Ability to model a large variety of testing strategies (including 
		pooled testing strategies) based on test availability, test cost, and 
		test accuracy and sensitivity;
	\item Option for a user to specify a custom stochastic viral load 
		profile which is used to determine an exposed agent's trajectory within 
		the model\footnote{An exposed agent is defined to be infected but not yet infectious.};
	\item Separate vaccinated and unvaccinated susceptible states;
	\item Ability to separately track isolated agents based on whether they 
		were isolated due to a false or true positive infection test;
	\item Ability to model loss of immunity by recovered agents; and
	\item Flexible agent trajectories (for example, an infectious agent does not 
		need to be isolated before it can become recovered).
\end{itemize}

This paper is organized as follows: \textit{Related work} expands on current 
designs in high-frequency and pooled testing, with primary emphasis given to 
work that combines both or acknowledges real-world limitations and complications 
of applied testing. Next, \textit{Model design} elaborates on choices made in the 
model design, involving the disease model and propagation, implementation of 
testing procedures, the impact of immunization, and cost estimation. Then, 
\textit{Experimental setup} and \textit{Results} demonstrate the flexibility and 
utility of \modelabrev{} by examining validation sims and their impact on model 
performance for multiple specific infectious disease transmission scenarios and 
corresponding mitigation strategies. We demonstrate reduced infection with 
more-rapid testing or test turnaround, diminished rates of false positives (and 
thus false isolation of healthy individuals) with pooled testing, and the 
robustness of these results under varied vaccination regimes. Finally, we 
conclude with suggested applications of \modelabrev{} and potential avenues for 
future improvement.


\section{Related work}
\label{sec:related_work}

In the monitoring of COVID-19, molecular assays are an important tool for detecting symptomatic and asymptomatic 
infections and have played a vital role~\cite{babiker,krause}. 
Quantitative real-time polymerase chain reaction (qRT-PCR) testing has been the gold 
standard for clinical diagnosis, due to the high sensitivity 
and specificity, but is expensive, requires highly-trained technicians, and incurs 
longer turn-around times \cite{Guglielmi_2020}. The expense and skilled workers 
required for qRT-PCR testing inhibits its application in resource-limited settings and in 
large-scale population screening. The delay between testing and reporting allows 
presymptomatic or asymptomatic individuals to spread COVID-19 prior to isolation. 
Because of these difficulties, several studies have suggested methods for reducing 
cost or improving response time of testing \cite{paltiel_2020,shental_2020,
augenblick_2022,lyng_2021,larremore_2021,nash_2021}.  

Suggestions for population-level screening have followed two prongs: high-frequency 
testing using low-cost tests \cite{larremore_2021,nash_2021} or pooled testing to increase the cost ratio 
of qRT-PCR tests \cite{shental_2020,augenblick_2022}. A problem with antigen tests 
is lowered sensitivity \cite{commissioner}, however, this is made up for by vastly 
reduced response time \cite{larremore_2021}. Pooled testing maintains the sensitivity 
of qRT-PCR, and increases the efficiency of testing in low-endemic scenarios 
\cite{augenblick_2022}, but is often accompanied by complicated pooling designs 
intended to optimize the one-shot throughput of testing \cite{shental_2020,hahn-klimroth_2022}. 
Two groups in particular \cite{augenblick_2022,lyng_2021} have attempted to marry 
these two prongs\textemdash{}combining pooled testing with higher-frequency test application 
to combat costs and result turn-around time. While no method has been superior to all 
others under all constraints, this is a promising direction that combines the 
strengths of both methods and provides an opportunity to adapt testing protocols 
as infection rates fluctuate \cite{cleary_2021,jonnerby_2020}. 

Continued testing of large cohorts, such as schools and businesses, for 
surveillance and prevention of pandemics is complicated and potentially prohibitively 
expensive. Testing regimens must be designed to minimize spread and reduce infection 
rates while also being simple and cheap enough to maintain for weeks or months 
on end \cite{science_2020,corman_2020}. Towards this end, several studies have 
explored methods for curbing COVID-19 spread in the face of social re-openings 
\cite{asgary_2021,lyng_2021,nash_2021,augenblick_2022,paltiel_2020}.

Most simulation studies have focused on one of two options for disease 
transmission mitigation: high-frequency testing or pooled testing. Proponents of 
high-frequency testing \cite{larremore_2021,paltiel_2020,grassly_2020,chin_2020} 
advocate for the distribution of antigen-based self-testing methods. These tests 
have lower sensitivity and specificity than qRT-PCR tests, but are significantly 
cheaper and have a turn-around time of minutes \cite{thi_2020,butler_2021,Meyerson_2020}. 
Proponents of pooled testing \cite{shental_2020,hahn-klimroth_2022,salcedo_2021,aldridge_2021,lock_2022} 
devise methods to use the sensitivity of qRT-PCR 
as an advantage, increasing the efficiency of individual tests. These studies 
predominantly implement two-stage Dorfman testing \cite{dorfman_1943} and generate complicated 
pooling designs. However, both options have shortcomings: the lack of sensitivity 
in antigen tests reduces their detection of asymptomatic or presymptomatic individuals, 
while complicated pooling designs are hard to adhere to in practice and ignore 
the repetitive nature of testing (frequent testing would make learning/implementing 
complicated pooling designs more robust).

Lyng \textit{et al.}~\cite{lyng_2021} and Augenblick 
\textit{et al.}~\cite{augenblick_2022} demonstrate the importance of modeling 
intervention strategies in conjunction, finding a hybrid high-frequency pooled 
testing approach to be more effective than either strategy alone. The 
latter~\cite{augenblick_2022} takes a theoretical approach, demonstrating 
enhanced efficiency in pooling designs through reduction in disease prevalence 
over time. While highly compelling, their simplified disease model and lack of 
tool make it hard to apply their results in practice. \
Lyng \textit{et al.}~\cite{lyng_2021} implement a stochastic, compartmental SIR 
disease model to explore pooling, frequency of testing, testing delays, as well 
as optimize for cost and sensitivity/specificity of tests. Both approaches 
demonstrate the benefits of a hybrid design, but suffer from simplistic disease 
models and time-invariant testing and pooling strategies.

\modelabrev{} extends these works \cite{lyng_2021,augenblick_2022} in several key 
ways. For a more complete simulation of the disease dynamics we extend the model 
to include other interventions, such as isolation and vaccination. This is 
reflected in a descriptive disease model, accounting for asymptomatic, 
presymptomatic, isolated, recovered, and imported infections using an 
agent-based model. Additionally, we implement custom viral load dynamics for all 
infected individuals, using methods from Cleary 
\textit{et al.}~\cite{cleary_2021} and Larremore 
\textit{et al.}~\cite{larremore_2021}. We maintain ideas from Lyng 
\textit{et al.}~\cite{lyng_2021} and explore the impact of test sensitivity, 
specificity, response time, and frequency. Finally, we provide all of this in a 
modular and computationally-efficient tool. These extensions provide users with 
the ability to simulate a much larger array of hybrid interventions on 
populations with a variety of characteristics. Additionally, the modularity 
allows expert users to adapt and extend our work to novel 
diseases~\cite{marani_intensity_2021} or population structures. The model's 
efficiency allows for users to simulate many disease control scenarios quickly 
and update them as new information becomes available.

The epidemiological model \modelabrev{} is based 
on\footnote{See Figure \ref{fig:disease-model}} is similar in concept to the 
Generalized SEIR model introduced by Liangrong Peng \textit{et al.} in 2020
and added to the MATLAB code base later that same year by E. Cheynet 
\cite{peng2020,cheynet2020}. Despite the apparent similarities, \modelabrev{} 
represents a significant extension of Generalized SEIR in terms of both 
flexibility and functionality. A summary comparison of \modelabrev{} with 
Generalized SEIR is provided in \ref{app:gSEIR}.


\section{Model design}
\label{sec:model_design}

\modelabrev{} is built on an agent-based model in which each agent is assigned a set of individual parameters to account for diversity in the population. This type of model provides flexibility to vary transitions between states based on individual characteristics such as vaccination status, viral load, and likelihood of self-isolation. States that agents can occupy are based on an enhanced SIR compartmental model. This set-up allows for the easy removal or addition of modules or compartments based on scenario characteristics. The currently implemented model includes modules to simulate testing, isolation, vaccination, and disease progression in terms of viral load and status of symptoms. All parameters associated with the various modules listed below are included in~\ref{app:params}.  

\subsection{Epidemiological model} \label{ssec:compartmental}

\begin{figure}[H]
    \centering
    \includegraphics[width=0.9\textwidth]{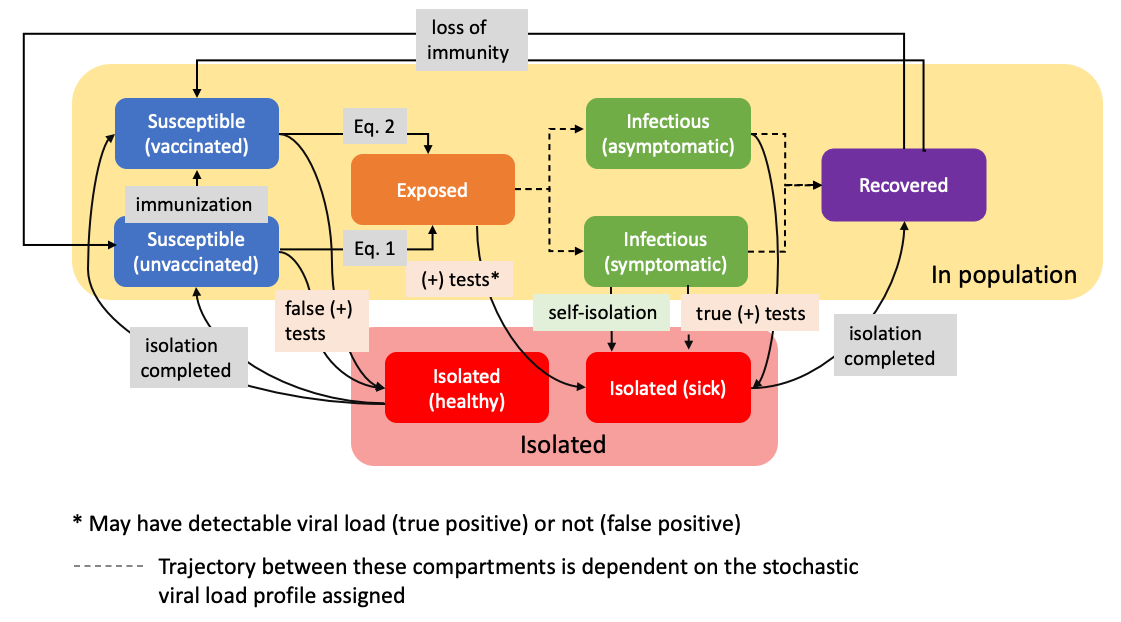}
    \caption{\textbf{Compartmental model.} Agents move between six compartments in the population (susceptible, exposed, infectious and recovered) and two compartments removed from the population (isolated while healthy or sick). The probability that a susceptible individual is exposed to the disease each day is given by Eq.~\ref{eq:su-exp} or Eq.~\ref{eq:sv-exp}, depending on their vaccination status. Once exposed, an individual's disease trajectory is determined by their viral load profile. A positive test at any point places an agent in isolation with ``healthy'' or ``sick'' status determined by their infection status. Additionally, agents who are experiencing symptoms may choose to isolate based on their pre-designated propensity to isolation on symptoms.}
    \label{fig:disease-model}
\end{figure}

Agents in the simulation move between six disjoint compartments following a variation of the SIR disease propagation model. We distinguish between when an agent is \textit{within} the population versus when it is \textit{isolated} from the population for easier computation of the disease propagation. Possible states within the population include: 
\begin{itemize}
    \item \textbf{Susceptible (vaccinated, $S_v$, or unvaccinated, $S_u$):} A \textit{susceptible} agent has the ability to be infected. The number of susceptible agents is given by the sum of unvaccinated susceptible and vaccinated susceptible, $S=S_v+S_u$.
    \item \textbf{Exposed, $E$:} An agent is in the \textit{exposed} category if it is infected but not yet infectious (based on a preset viral load infectiousness threshold). 
    \item \textbf{Infectious (symptomatic, $I_s$, or asymptomatic, $I_a$):} An \textit{exposed} agent becomes a \textit{infectious} agent once its viral load surpasses the designated infectiousness threshold. An infectious agent can be either symptomatic or asymptomatic, $I=I_s+I_a$.   
    \item \textbf{Recovered, $R$:} An agent has \textit{recovered} once its viral load falls below the infectiousness threshold. 
\end{itemize}
The total number of agents in the population (not in isolation) at a given time is given by $P = S + E + I + R$. Additionally, if an agent is isolated it is in either the \textit{isolated (healthy)} or \textit{isolated (sick)} state, where the \textit{isolated (healthy)} state is comprised of agents who received a false positive test result.

\subsubsection{Exposure} \label{ssec:exposure}
The probability that an unvaccinated susceptible agent is exposed to the disease on a given day is given by the sum of the probability an agent is exposed outside the population and the mass action probability of being exposed inside the population: 
\begin{align}
    \mathbb{P}(S_u \rightarrow E) = \gamma + \beta \frac{I}{P},
    \label{eq:su-exp}
\end{align}
where $\gamma$ is the probability of being infected outside the population and $\beta$ is the typical interaction parameter. We assume a well-mixed population and randomly choose $\mathbb{P}(S_u \rightarrow I) \cdot S_u$ of the $S_u$ susceptible agents to be exposed.

The exposure process for vaccinated susceptible agents is similar to~\eqref{eq:su-exp}, with the addition of a immunity discount factor, $\alpha$: 
\begin{align}
    \mathbb{P}(S_v \rightarrow E) = \alpha \left(\gamma + \beta \frac{I}{P}\right).
    \label{eq:sv-exp}
\end{align}
As in the previous case, $\mathbb{P}(S_v \rightarrow E) \cdot S_v$ are randomly chosen from $S_v$ to become exposed. 

\subsubsection{Viral load progression} \label{sub:vlm}
The progression of disease transmissibility and symptoms is characterized by the 
disease and can be highly variable between individuals~\cite{he_temporal_2020}. 
For the COVID-19 based scenarios explored in Sections 4 and 5, we demonstrate 
\modelabrev{}'s ability to utilize a user specified viral load evolution model by 
implementing one based on the work of Larremore 
\textit{et al.}~\cite{larremore_2021}. This models the viral load as having a 
hinge-function profile (consistent with Marc 
\textit{et al.}~\cite{marc_quantifying_2021}) with variations between 
asymptomatic and symptomatic individuals. 

At the time of exposure, the newly exposed individuals are chosen to be symptomatic or asymptomatic with probability $\sigma_s$ (\ref{app:diseaseparams}: \texttt{fractionSymptomatic}). An agent is assigned a set of viral load parameters chosen from the corresponding distributions (described below and summarized in Table~\ref{table:viral-load}). The resulting viral load progression influences an agent's progression from \textit{exposed} $\rightarrow$ \textit{infectious (symptomatic or asymptomatic)} $\rightarrow$ \textit{recovered}. Additionally, an agent's viral load directly affects the results of any testing that may take place during this period. The structure of this module and distribution of parameters can be modified to model the progression of a different disease. 

\begin{table} 
    \centering
    \begin{tabular}{|c|c|p{0.35\linewidth}|}
        \hline 
        Parameter & Distribution & Description\\ \hline \hline
        $\sigma_s$ & 0.5 & Probability of an agent being symptomatic \\ \hline
        $t_0$ & $\text{uniform}(2.5, 3.5)$ & Time interval of viral load initialization  \\ \hline
        $V_0$ & $10^3$ cp/ml & Initial viral load \\ \hline
        $t_P$ & $\Gamma(1.5,1) + 0.5$ & Time interval to achieve peak viral load \\ \hline 
        $V_P$ & $\text{uniform}(10^4,10^7)$ & Peak viral load \\ \hline
        $t_S$ & $\text{uniform}(0,3)$ & Time interval for symptoms to begin \\ \hline
        $t_F$ & $\text{uniform}(4,9)$ & Time interval for viral load to decline to $V_F$ level\\ \hline
        $V_F$ & $10^3$ cp/ml & Final viral load level\\ \hline
        $V_I$ & $10^3$ cp/ml & Minimum viral load for infectiousness\\ \hline
    \end{tabular}
    \caption{\textbf{Viral load parameters and their corresponding distributions.} }
    \label{table:viral-load}
\end{table}

\begin{figure}[H]
    \centering
    \begin{subfigure}[h]{0.47\textwidth}
        \includegraphics[width=\textwidth]{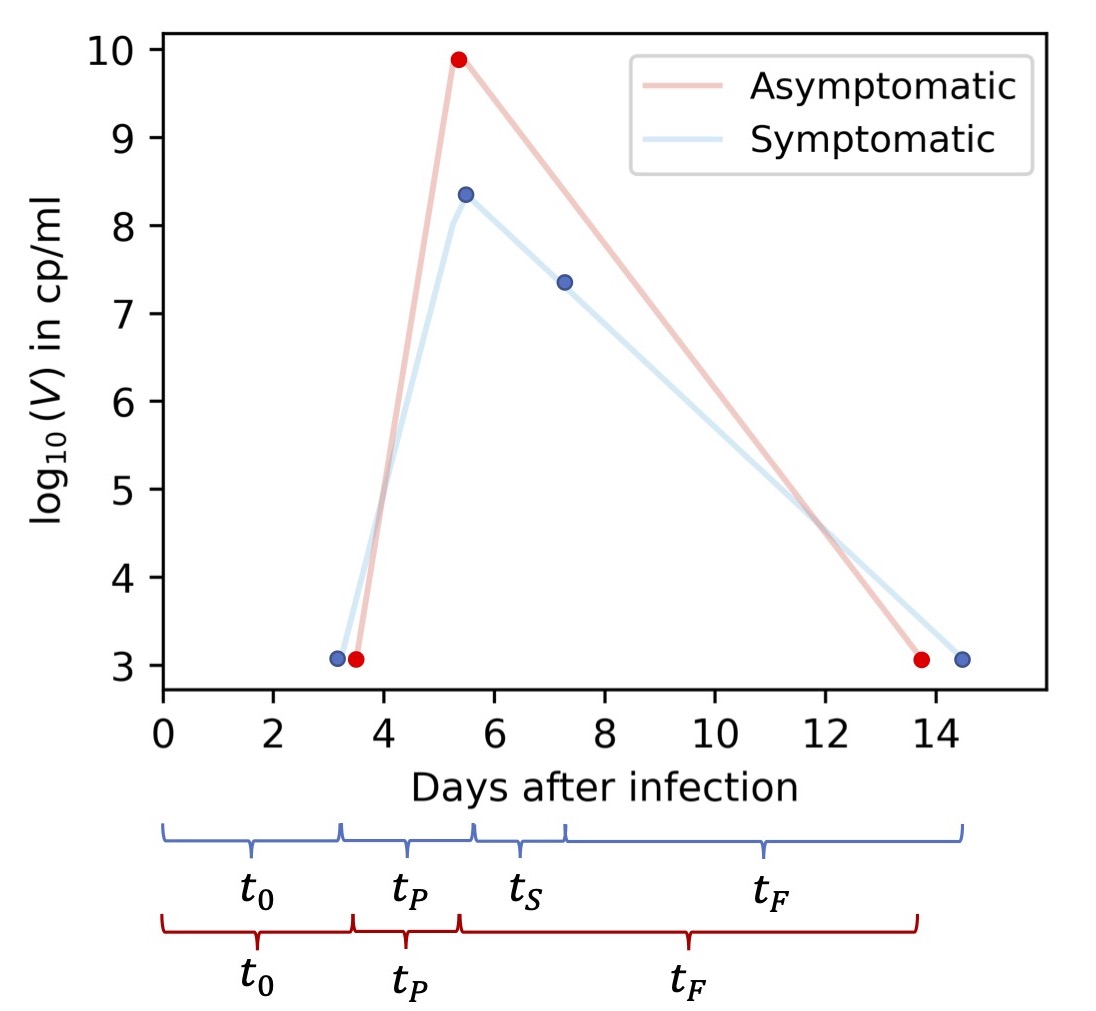}
        \subcaption{\textbf{Viral load model.}}
        \label{fig:viral-load-params}
    \end{subfigure}    
    \hfill
    \begin{subfigure}[h]{0.47\textwidth}
        \includegraphics[width=\textwidth]{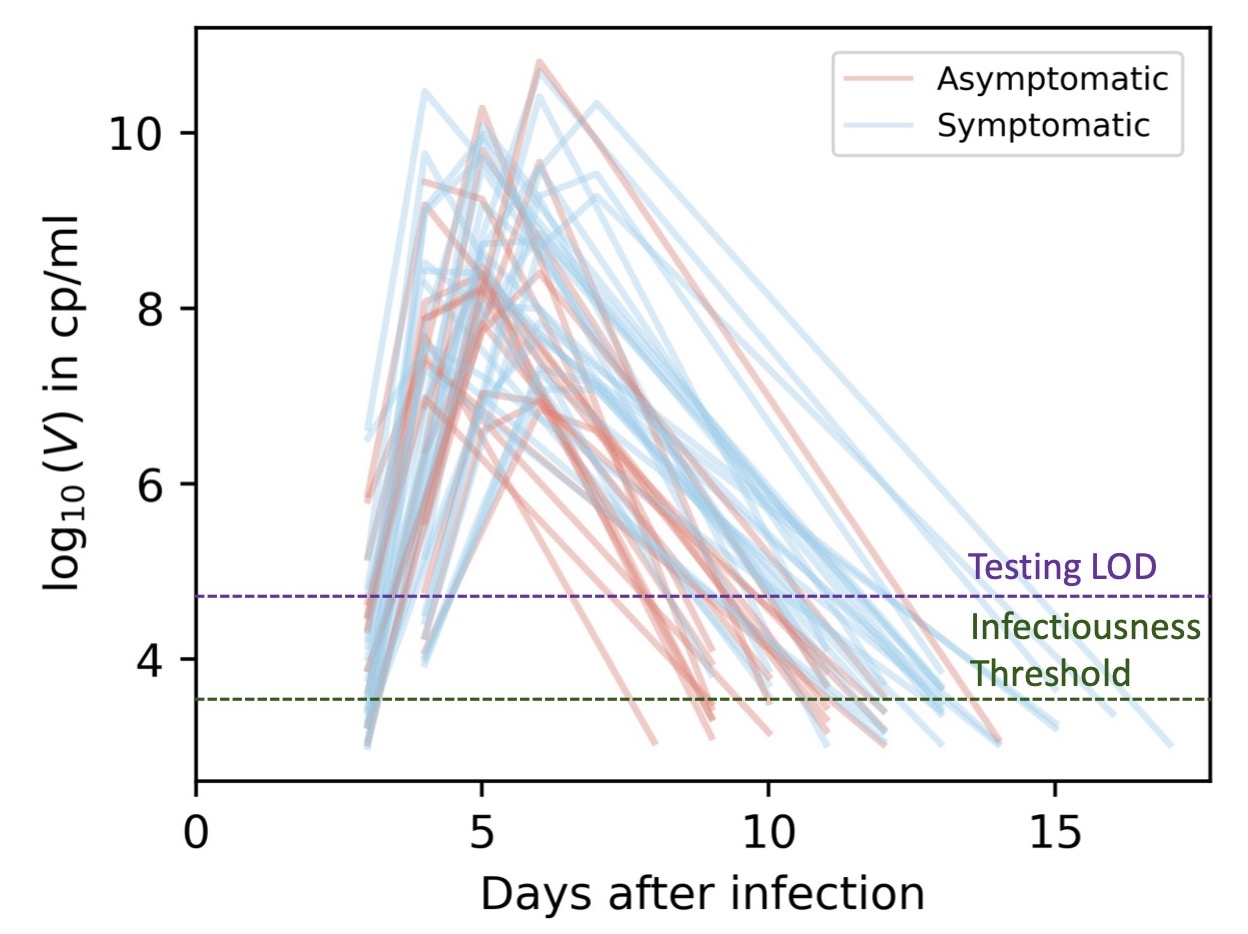}
        \subcaption{\textbf{Example viral load profiles.}}
        \label{fig:viral-load}
    \end{subfigure}    
    \caption{\textbf{Viral load model.} Description and example trajectories associated with the viral load model based on that of Larremore \textit{et al.}~\cite{larremore_2021}. Note that although panel~\ref{fig:viral-load-params} shows the asymptomatic trajectory achieving a higher viral load than the symptomatic trajectory, in general this is not the case. These are two sample trajectories from the stochastic process based on the distributions in Table~\ref{table:viral-load}.}
    \label{fig:viral-load-full}
\end{figure}

The asymptomatic viral-load trajectory is a hinge-function defined by the (time (days), viral load (cp/ml)) coordinates:
\begin{align}
    (t_0,V_0) \rightarrow (t_0+t_P,V_P) \rightarrow (t_0+t_P+t_F,V_F), 
\end{align} 
where each variable is drawn from the distributions in Table~\ref{table:viral-load}. The trajectory for a symptomatic individual is similar, with the addition of the appearance of symptoms $t_S$ days after achieving peak viral load. This also results in a prolonged decrease of viral load back to the initial baseline. The function coordinates are:  
\begin{align}
    (t_0,V_0) \rightarrow (t_0+t_P,V_P) \rightarrow (t_0+t_P+t_S+t_F,V_F).
\end{align} 
This process along with fifty resulting viral load profiles is shown in Fig.~\ref{fig:viral-load-full}.

Once an agent's viral load reaches a user designated threshold for infectiousness (\ref{app:diseaseparams}:\;\texttt{infectiousViralLoadCut} ($V_I$)) it is moved from \textit{exposed} to either \textit{infectious (symptomatic)} or \textit{infectious (asymptomatic)}. Similarly, once an infectious agent's viral load drops below the infectiousness threshold, the agent is moved to \textit{recovered}. If $V_0=V_F=V_I$ the total time an asymptomatic (symptomatic) agent is infectious is $t_0+t_P+t_F$ ($t_0+t_P+t_S+t_F$).  

\subsection{Population interventions}

\subsubsection{Testing}
One of the most common types of interventions for infectious disease transmission mitigation is population testing. The numerous types of tests, schedules, and costs involved make this a key process for decision makers to optimize. These types of decisions often have multiple objectives as employers wish to limit the spread of disease while also limiting the cost and mental health impact associated with unnecessary isolation resulting from false positive tests. Our simulation is able to quickly compare many scenarios and can be used to inform disease control decisions. 

We offer users the ability to specify parameters for test schedule (initial day of testing and testing frequency) and optionally for two-stage Dorfman pooling (pool size and function to use for determining pooled test outcome). Different types of tests can also be compared by setting the viral load threshold at which the disease can be detected, as well as a false negative rate, false positive rate, and delay for the return of test results. See the full list of testing parameters in~\ref{app:testingparams}.

When testing is performed, all eligible individuals are split into pools of the designated size. If a pool consists of a single individual, then single testing is performed, otherwise we use a two-stage Dorfman pooled testing procedure~\cite{dorfman_1943}. 

In the single sample testing procedure, each individual $i$ is considered \textit{detectable} if their viral load ($v_i$) is greater than test $m$'s limit of detection ($l_m$). The single sample testing result $t_m(i)$ of sample $i$ is then positive with probability: 
\begin{align}
    \mathbb{P}\big(t_m(i)=+\big) = \begin{cases} 
        \phi_p & v_i \leq l \\
        1-\phi_n & v_i > l
    \end{cases},
\end{align}
where $\phi_p$ and $\phi_n$ are the false positive and false negative rates associated with the test. 

The two-stage Dorfman pooled testing procedure is similar to that of the single testing procedure, but proceeded by a test applied to each pool. We offer users two functions for determining the pooled test results, average pooling and exponential pooling. Both are defined in~\ref{app:testingparams}:\texttt{poolingType}, but we limit the discussion here to the default option, average pooling. We use an apostrophe to distinguish the viral load and testing result of a pool ($v'_j$, $t'(j)$) from that of a single sample ($v_i$, $t(i)$).

In this case, the viral load content of a pool $j$ is defined as the average of the viral load of all samples in the pool. That is, 
\begin{align}
    v'_j = \displaystyle\sum_{i=1}^N v_i/N,
\end{align}
where $N$ is the size of the pool. As above, a pool is considered detectable if the viral load, $v'_j$, is greater than the test's limit of detection. The probability that pool $j$'s test result, $t'_m(j)$, is positive is,
\begin{align}
    \mathbb{P}\big(t'_m(j)=+\big) = \begin{cases} 
        \phi_p & v'_j \leq l \\
        1-\phi_n & v'_j > l
    \end{cases}.
\end{align}
After testing each pool (stage 1), the second stage consists of applying single sample testing to every member of each positive pool.

\subsubsection{Isolation}
Isolation procedures and self-isolation play a key role in removing infectious individuals from a population. We consider two cases of isolation: isolation due to receipt of positive testing results or self-isolation due to experiencing symptoms. We assume agents in the first case are entirely compliant as this may be enforceable by an employer, testing official, etc. Isolation parameters in this simulation include length of isolation and the probability that agents self-isolate when symptoms are experienced. Additionally, users may choose to enact the withholding of tests for a set period of time after agents have exited isolation and recovered. The full list of isolation parameters are included in~\ref{app:isolationparams}. 

In the case of an agent testing positive, if they are exposed or infectious at the time of testing positive, they are moved to the \textit{isolation (sick)} compartment. If the agent is susceptible at the time of testing (i.e. they tested positive falsely), they are moved to the \textit{isolation (healthy)} compartment. Additionally, when an agent experiences symptoms, they may choose to self-isolate based on the preset probability of self-isolation. These individuals are also moved to the \textit{isolation (sick)} compartment. 

After a set number of days, agents in the \textit{isolation (sick)} compartment are moved to \textit{recovered}. Likewise, individuals in the \textit{isolation (healthy)} compartment are moved back to susceptible.

\subsubsection{Vaccination}\label{ssec:vaccination}
Once a vaccine has been developed for an infectious disease, immunization of a population is one of the most effective methods of reducing the impact of an infectious disease. In this simulation users are able to simulate the distribution of vaccines by specifying the rate at which vaccines are available to the population. Agents' preferences can also be modeled by specifying a distribution of vaccine acceptance among agents. In this case agent $i$ is assigned a ``willingness to vaccinate'' probability, $\nu_i$, between 0 and 1 drawn from the distribution. At each time step, $t$, agents are labeled as ``willing'' to vaccinate with probability: 
\begin{align}
    \mathbb{P}(\text{agent $i$ is willing to vaccinate during time step $t$}) = \nu_i.
\end{align}
The vaccines that are available are distributed to the willing agents until all willing agents have been vaccinated. See~\ref{app:vaxparams} for the full list of vaccination parameters.

\subsection{Ordering of simulation procedures}
All simulation processes are described above in detail, but for completeness, we also include here the order in which each of these processes takes place during a single simulation time step or ``day.'' 

\begin{enumerate}
    \item \textbf{External exposure}: The first stage of the simulation records exposure that occurred outside the population (the first term in Eqs.~\ref{eq:su-exp} and~\ref{eq:sv-exp}). Exposed agents are labeled as symptomatic or asymptomatic and assigned a set of viral load parameters according to Section~\ref{sub:vlm}. At this point the full viral load timeline is also saved for easy reference throughout the simulation.   
    \item \textbf{Update agent status}: This stage encompasses the bulk of agent movement between compartments and parameter updates. This includes: 
    \begin{itemize}
        \item Advancing the viral load of each exposed and infectious agent by one time step 
        \item Movement of agents between compartments based on any testing results received on this day 
        \item Agents' exit from isolation if they have completed the designated number of days 
        \item Movement from \textit{exposed} to \textit{infectious} for any agents with viral load above the infectiousness threshold 
        \item Movement from \textit{recovered} to \textit{susceptible} based on the number of days since each agent's recovery 
    \end{itemize}
    \item \textbf{Self-isolation}: A subset of the agents which are symptomatic are placed into isolation based on the propensity to self-isolate parameter.  
    \item \textbf{Testing}: If the current time step is designated as a testing day, samples are pooled (if applicable) and testing is performed. Results are scheduled to be received in the future based on the parameter delaying test results.  
    \item \textbf{Infection propagation}: Agents are infected based on exposure in the population (second term of Eqs.~\ref{eq:su-exp} and~\ref{eq:sv-exp}). Again, agents are marked as symptomatic or asymptomatic and assigned a viral load timeline.  
    \item \textbf{Vaccination}: Eligible agents (not in isolation and not yet vaccinated) are chosen for vaccination based on the number of available vaccines and each agent's propensity to vaccinate.  
\end{enumerate}

\subsection{Implementation}
\modelabrev{} is implemented in Python and features scripts for duplicating results from this paper as well as creating new disease scenarios. Many parameters are dependent on the disease and population being modeled, and thus we leave their selection to the researchers and decision-makers with knowledge of the specifics. However, some general strategies for estimating population parameters such as ``propensity to isolate'' may be to provide a survey to employees or estimate from the general public. The simulations performed in Section~\ref{sec:experiment} took around 6 seconds per scenario using a single CPU.


\section{Experimental setup}
\label{sec:experiment}

Capabilities of \modelabrev{} were demonstrated through a series of simulations. Our goal was to showcase scenarios where the tool could be used to evaluate alternative courses of action for management of a disease in a population. Disease dynamics for a simulated population of 10,000 agents were examined for a variety of vaccination and testing scenarios. To our knowledge, no non-healthcare business 
provided vaccines for COVID-19, and thus our simulations assume an exogenous application 
and uptake of vaccines in the population. Three different vaccination scenarios (Table~\ref{table:vax}) loosely represent dynamics in a population without any vaccination (Vaccination A), during vaccine rollout in an unvaccinated population (Vaccination B), or during continued vaccine distribution in a partially vaccinated population (Vaccination C). 

Within each of these vaccination settings the effect of several testing schemes were evaluated. There are several types of tests available, thus it is within scope to assume that management would be considering which test to employ. For our simulations, two types of tests (Table~\ref{table:test}) were considered. Test A had similar characteristics to a PCR test, with higher sensitivity but a longer turnaround time for results. Test B was more similar to an antigen test with lower sensitivity and shorter turnaround time. False positive and false negative rates for the tests were approximated by averaging over a subset of approved PCR and antigen tests~\cite{molecular_fda_2023, antigen_fda_2023}. Example limits of detection (LOD) in number of genetic copies per milliliter (cp/mL) for each type of test were based on~\cite{cubas-atienzar_limit_2021,arnaout_sars-cov2_2020} and verified with~\cite{molecular_fda_2023, antigen_fda_2023}. 

For each vaccination scenario and test combination, simulations were run for different testing intervals (4 or 7 days) and pooled testing scenarios (5 sample pooling, no pooling). All other parameters 
(\ref{app:params})
were held constant. Disease parameter $\beta$ was derived following~\ref{app:beta}. Each unique simulation configuration was run 50 times to demonstrate consistency between runs. 

\begin{table}[H]
    \centering
    \begin{tabular}{|c||c|c|c|}
        \hline 
       Parameter & Vaccination A & Vaccination B & Vaccination C \\
        \hline\hline
        {\tt initProportionVaccinated} & 0 & 0 & 0.5 \\
        \hline 
        {\tt vaccinesAvailablePerDay} &  0 & 50 & 50 \\
        \hline 
    \end{tabular}
    \caption{\textbf{Vaccination scenarios.} Three different vaccination scenarios were created to represent a population without any vaccination (Vaccination A), a population of unvaccinated individuals with some vaccine distribution (Vaccination B), and a partially-vaccinated population with vaccine distribution (Vaccination C). }
    \label{table:vax}
\end{table}

\begin{table}[H]
    \centering
    \begin{tabular}{|c||c|c|}
        \hline 
       Parameter & Test A & Test B  \\
        \hline\hline
        {\tt fprSingle} & 0.014 & 0.007  \\
        \hline 
        {\tt fnrSingle} &  0.06 & 0.15  \\
        \hline
        {\tt detectionCut} &  100 cp/ml & $1.0 \times 10^6$ cp/ml  \\
        \hline
        {\tt daysDelayTestResults} &  3 & 0  \\
        \hline 
        {\tt cost} &  \$100 & \$50  \\
        \hline 
    \end{tabular}
    \caption{\textbf{Testing scenarios.} Test A was modeled after a PCR test with sensitivity, specificity, and LOD estimated from~\cite{molecular_fda_2023} and~\cite{arnaout_sars-cov2_2020}. Test B was modeled after an antigen test with values estimated from~\cite{antigen_fda_2023, cubas-atienzar_limit_2021}. We also assume test A is more expensive with a per test cost of \$100 compared to a cost of \$50 for each application of test B. These values are used to compare the total expense of implementing each testing scenario. Simulations were performed with either Test A or Test B and evaluated for testing intervals ({\tt daysBetweenTesting}) of 4 or 7 days and pool size ({\tt poolSize}) of 1 or 5 samples.}
    \label{table:test}
\end{table}

\clearpage


\section{Results}
\label{sec:results}

We examine the simulated scenarios based on goals from a small-company or 
managerial perspective: maximal safety for our employees, as reflected by minimizing 
the total number of infections, minimal loss of effective time, as reflected 
by minimizing the number of falsely isolated individuals (e.g. healthy people placed in isolation), 
and minimal expense to the company. We first show results for a population without vaccination (Vaccination A) while discussing how they translate to the other vaccination scenarios (results included in~\ref{app:resultsBC}).

In order to visually demonstrate how testing choices impact the main compartments of interest, we directly compare the cumulative number of people who enter the exposed category (total infections) or are falsely isolated. We also compare the total cost of each scenario assuming a per test cost of \$100 for test A and \$50 for test B. More detailed disease dynamics showing the number of people occupying each compartment of the model (Fig.~\ref{fig:disease-model}) over time can be found in~\ref{app:resultsA}. 

\subsection{Reducing testing interval is the most effective way to reduce 
infections} \label{sub:inf_trans}

The most important contribution of our tool is reduced harm to employees. We first 
demonstrate this in Fig.~\ref{fig:vaxA} in a population without vaccination (Vaccination A), 
while exploring the impact of pooling and/or reduced 
testing interval. Looking first at the application of test A, we see a decreased number of infections (orange) when testing every 4 days compared to weekly testing. It should be noted that this also comes with an increased cost, although that can be mitigated with pooling of samples (Section~\ref{sub:test_Admin}).

For comparison, under the same vaccination status 
but a faster test return (test B, Fig.~\ref{fig:vaxA}), we see a similar reduction in the number of infected people between weekly testing scenarios and more frequent testing. In fact, even though test B has a higher rate of false negative, it is more effective at reducing infection (and cost) than test A in most scenarios compared due to its faster turnaround time. The most effective scenarios among test types and testing intervals involve both a faster test turnaround and more frequent testing. 

These differences demonstrate that among the options explored, more frequent testing is the most consistent method for reducing disease 
incidence in a population. Additionally, using a test with faster 
turn-around can further add to these effects since results are known more quickly for rapid isolation of people who may transmit disease.

\subsection{Pooled testing is more effective at reducing false isolation
than testing interval} \label{sub:false_q}

After ensuring the safety of workers, a company's next concern is loss of productivity, 
indicated by total people hours lost to infections. Reduced unnecessary isolation 
also reduces the social impact of a pandemic on a society, improving the mental health of citizens and possibly increasing 
adherence to testing and quarantine regimes. Therefore, it is not only in a single 
business' best interest to reduce false isolation, but there are additional 
benefits to society outside of work.  

\begin{figure}
    \centering
    \includegraphics[width=\textwidth]{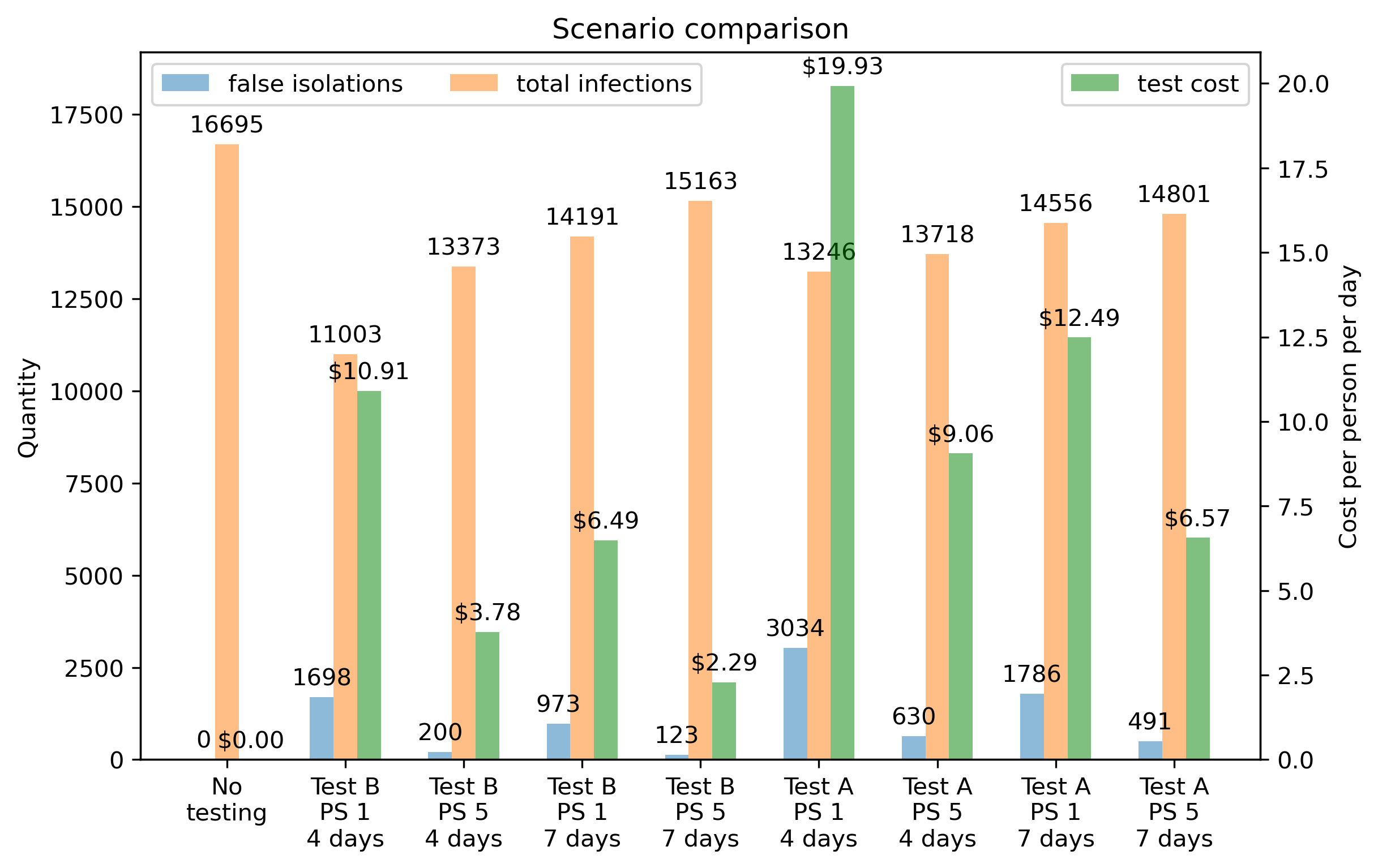}
    \caption{\textbf{Comparison of testing scenarios in a population with no vaccination (Vaccination A)}. Scenario labels correspond to the type of test used, pool size (`PS') used for pooled testing, and testing interval, respectively. Left and center bars (number of false isolations and total infections) correspond to the y-axis on the left, while the right-most bars (test cost) correspond to the y-axis on the right. Testing scenario cost is based on a cost of \$100 for test A and \$50 test B. Total testing costs are divided by 120 days and 10,000 people in the population to get the cost per person per day.}
    \label{fig:vaxA}
\end{figure}

Figure~\ref{fig:vaxA} (false isolations, `PS 1', `4 days' vs. `7 days'), demonstrates the impact of testing 
interval on false isolation. Paradoxically, longer time between testing reduces 
the amount of healthy people isolated. This is an artifact of single tests, where 
the false-positive rate is independent for each test, and thus testing more results 
in increased false isolation. 

In contrast, pooled testing provides a significant reduction in false isolation. Pooling tests reduces the 
false-positive rate of a pool through our two-stage application - the probability 
of two false positives is negligible. As such, the minimization of false isolation 
holds across longer testing regimes (Fig.~\ref{fig:vaxA}, `7 days') and even with reduced test efficacy (Fig.~\ref{fig:vaxA}, `Test B').

All of these results are directly compared in Fig.~\ref{fig:vaxA}. 
Here, we see the importance of pooled testing, compared to individual testing, 
under all test efficacy and interval combinations that we explored. Test efficacy is more 
important than test interval, as test A scenarios generated significant false 
isolation compared to test B, but this effect is reduced when pooled testing is used. 
This demonstrates the large reduction in false isolation provided by pooled testing.

\subsection{Test cost is effectively reduced under pooling regimes} 
\label{sub:test_Admin}

To be effective, companies must be aware of the bottom line. While ensuring the 
safety of our employees, it is important to acknowledge the costs of our decisions 
and, if possible in a safe and effective manner, reduce the expenditure of those 
actions. Our tool allows direct control over costs by allowing different tests to 
be provided to simulations. Additionally, a more indirect (but more effective) cost reduction 
is the application of fewer tests. While this cannot be planned \textit{a priori}, 
we can explore which scenarios provide the greatest reduction in test usage as incorporated into reduced overall cost. 

Figure~\ref{fig:vaxA} shows the total cost per person per day for the tests administered for each scenario. 
First, we find the obvious conclusion - increasing the testing interval reduces the 
cost of tests provided. However, we strongly recommend not taking this option, as 
previous sections have shown reduced testing to increase the incidence of disease 
in the population and have small benefits for reducing the amount of false isolation. 

However, pooled testing also has a significant impact on the total cost of a testing strategy, even more than increasing the testing interval. This option has 
also been shown to significantly reduce the number of 
healthy people put into isolation. As such, we believe that pooled testing demonstrates 
the safest method for reducing the testing burden on a company.

\subsection{The impacts of reduced test interval and pooled testing hold across 
vaccination regimes} \label{sub:vax_gen}

We are no longer at the initial stages of the COVID-19 pandemic. At some point, there 
may be another pandemic, where we need surveillance of a naive population \cite{marani_intensity_2021}. However, 
most nations have begun vaccinating their populations and we are now at a state 
of partial vaccination, with continued vaccine rollout, while we continue 
surveillance testing. Therefore, we have built our tool to integrate the current 
levels of employee vaccination within a company, as well as continued vaccination 
of employees, and we explored the impact of partial vaccination on testing regimes 
for improved incidence reduction (Table~\ref{table:vax}). The figures from vaccination scenarios 
B and C can be found in~\ref{app:resultsBC}.  

It should be noted that overall, vaccination of the population does a more effective job at effectively reducing total infection than testing alone (Fig.~\ref{fig:vaxA} vs. Figs.~\ref{fig:vaxB} and~\ref{fig:vaxC}). That being said, the results outlined in previous sections hold - the impacts of short versus long 
intervals between tests on incidence, individual vs pooled testing on false isolation, 
and across test efficacy. Although the absolute differences in testing regimes appears reduced 
due to the reduced size of the susceptible population, this does not diminish 
the trends for reducing infectious cases or false isolation. In fact, it improves 
the absolute impact of our testing regime, further reducing the amount of infected 
people or healthy people sent unnecessarily into isolation.

Interestingly, the effect of pooling on test expense is actually \textit{increased} in a population with higher levels of vaccination. The cost per person associated with single testing increases with the size of the susceptible population, while the cost of pooled testing decreases based on the corresponding lower rate of incidence (test cost, Fig.~\ref{fig:vaxA} vs. Figs.~\ref{fig:vaxB} and~\ref{fig:vaxC}). These observations demonstrate 
the utility of our tool at the early, middle, and through the later stages of a 
pandemic.


\section{Conclusion}
\label{sec:conclusion}

The COVID-19 pandemic spurred a need for creation and implementation of testing protocols, quarantine guidelines, and vaccine distribution. Knowledge gleaned from COVID-19 informed the creation of our disease management tool for rapid comparison of infection propagation and response scenarios. While management tools were limited at the beginning of the COVID-19 pandemic, the flexibility of this simulation allows for easy adaptation to different disease scenarios resulting in a shorter start-up time for future epidemics. 

The ability to quickly compare the impact of disease interventions is invaluable in management of an epidemic. By integrating infection propagation, agent isolation, testing, and vaccination, this tool can provide non-trivial insights into the relative effectiveness of disease transmission mitigation strategies in a population. This work demonstrated the tool's utility in the context of COVID-19 management in several populations representing stages of a pandemic. We show that in each of these stages, simulation can be useful for determining the relative effects of management decisions. 

In the future, this work could be extended to account for the impact of relationships on disease propagation, with the implementation of social interaction networks \cite{pine2023}. This would provide the ability to simulate the effect of interventions such as varied work schedules. Additionally, we now know that there are several possible outcomes from COVID-19 infection, such as full recovery or partial recovery (``long-term COVID-19''). The model could be expanded to account for different recovery states, as well as death, as this is a more reliable metric for matching simulated outcomes against realized, real-world impacts.

\def\UrlBreaks{\do\/\do-}
\bibliographystyle{elsarticle-num-names-urldate} 
\bibliography{references-els}

\section*{Acknowledgements}
Work on this research has been funded by the Air Force Research Laboratory (AFRL) Autonomy 
Capability Team 3 (ACT3) under contracts FA8650-20-C-1121 (K.P., J.K.) and FA8649-20-C-0130 (R.V., J.B.). 
The authors would like to thank Dr. Michael Mendenhall, Dr. Jared Culbertson, and Dr. Scott Clouse for their assistance. 

\section*{Author contributions statement}
K.P., R.V., and J.K. conceived the idea for the simulation tool. K.P. and R.V. developed the code and ran simulations. K.P., J.B., and J.K. assisted in writing the manuscript and preparing for publication. 

\section*{Additional information}
\subsection*{Competing interests}
The authors declare no competing interests.

\newpage
\appendix
\setcounter{figure}{0}
\section{Simulation parameters} \label{app:params}
The full list of parameters for the \modelabrev{} are listed below by model category. The default values used for the experiments are included in parentheses (excepting those included in Tables~\ref{table:vax} and~\ref{table:test}). References linking these values to relevant research on the COVID-19 dynamics are included. 

\subsection{Run parameters}\label{app:runparams}
\begin{itemize}
    \item {\tt popSize} (10,000): Number of agents in the population
    \item {\tt timeHorizon} (120): Length of the simulation in days 
    \item {\tt initialInfected} (200 or 2\% of the population): Number of agents infected at the start of the simulation
    \item {\tt initProportionVaccinated}: Proportion of the population initially vaccinated
\end{itemize}

\subsection{Epidemiological model}\label{app:diseaseparams}
\begin{itemize}
    \item {\tt betaDaily} (0.4): Beta ($\beta$) parameter for daily infection propagation (Eqs.~\ref{eq:su-exp} and~\ref{eq:sv-exp})\footnote{See Appendix \ref{app:beta} for a discussion on estimating $\beta$.}
    \item {\tt daysTilSusceptible} (30): Number of days until an agent becomes susceptible after recovery~\cite{layton_understanding_2022}
    \item {\tt externalExposureProbDaily} (0.005): Daily probability ($\gamma$) of being exposed outside of the population
    \item {\tt fractionSymptomatic} (0.5): Proportion of infected individuals ($\sigma_s$) who develop symptoms~\cite{layton_understanding_2022,ma_global_2021}
    \item {\tt infectiousViralLoadCut} ($10^3$): Viral load, $V_I$, needed for an agent to become infectious (in cp/ml)~\cite{aspinall_viral_nodate,arnaout_sars-cov2_2020}
    \item Viral load model parameters~\cite{larremore_2021}
    \begin{itemize}
        \item {\tt t0} ($\text{uniform}(2.5, 3.5)$): Time interval of viral load initialization  
        \item {\tt V0} ($10^3$): Initial viral load (cp/ml)
        \item {\tt tP} ($\Gamma(1.5,1) + 0.5$): Time interval to achieve peak viral load
        \item {\tt VP} ($\text{uniform}(10^4,10^7)$): Peak viral load in cp/ml
        \item {\tt tS} ($\text{uniform}(0,3)$): Time interval for symptoms to begin 
        \item {\tt tF} ($\text{uniform}(4,9)$): Time interval for viral load to decline to $V_F$ level
        \item {\tt VF} ($10^3$): Final viral load level in cp/ml  
    \end{itemize}
\end{itemize}

\subsection{Testing}\label{app:testingparams}
\begin{itemize}
    \item {\tt daysBetweenTesting}: Interval at which testing is performed for the population
    \item {\tt daysDelayTestResults}: Number of days before a test result is received  
    \item {\tt detectionCut}: Viral load needed for detection by a test (in cp/ml)
    \item {\tt firstDayOfTesting} (7): First day of the simulation to perform testing 
    \item {\tt fprSingle}: False positive rate for a single test
    \item {\tt fnrSingle}: False negative rate for a single test
    \item {\tt poolingType} ({\tt average}): Function to use for computing pooled test results 
    \begin{itemize}
        \item {\tt average}: Pool detectability is determined by the average of the viral load of all samples in the pool (i.e. a pool is detectable by test $m$ if $\sum_{i=1}^N v_i/N > l_m$, where $v_i$ is the viral load of sample $i$, $N$ is the size of the pool, and $l_m$ is the limit of detection of test $m$)
        \item {\tt exponential}: Pools are assigned a false positive and false negative testing rate based on the number ($k$) of detectable samples in the pool. The pooling false positive rate is equivalent to that of a single test ($\phi_p$), since in either case the viral load is below the detectable threshold. The false negative rate of a pool is equal to the false negative rate of a single test raised to $k$ ($\phi_n^k$) since each detectable sample contributes to the viral load of the pool and thus decreases the chance of a negative test. In summary, the probability of a pool testing positive is:  
        \begin{align*}
        p_+ = \begin{cases} 
            \phi_p & k = 0 \\
            1-\phi_n^k & k > 0 
        \end{cases}.
        \end{align*}
    \end{itemize}
    \item {\tt poolSize}: Pool size for test processing (a value of 1 corresponds to no pooled testing)
\end{itemize}

\subsection{Isolation}\label{app:isolationparams}
\begin{itemize}
    \item {\tt noTestingPostIsolationDays} (0): Number of days to delay testing of an agent after release from isolation 
    \item {\tt isolationLength} (10): Number of days an agent is in isolation
    \item {\tt selfIsolationOnSymptomsProb} (0.7): Probability that an agent self-isolates when they experience symptoms 
\end{itemize}

\subsection{Vaccination}\label{app:vaxparams}
\begin{itemize}
    \item {\tt vaccineAcceptProbMean} (0.7): Mean probability that agents are willing to vaccinate 
    \item {\tt vaccineAcceptProbStd} (0.05): Standard deviation in distribution of probability that agents are willing to vaccinate 
    \item {\tt vaccinesAvailablePerDay}: Number of vaccines available for distribution during each day of the simulation 
    \item {\tt vaccineInfectionProb} (0.3): Probability of exposure ($\alpha$) for vaccinated agents  (average over the simulation duration based on values in~\cite{hall_protection_2022}) 
\end{itemize}

\section{Determining $\beta$} \label{app:beta}
One method for estimating the disease parameters associated with a novel infectious disease is to fit them to the \textit{basic reproductive number}, $R_0$, associated with early stages of an epidemic. For the simulations in this paper, the $\beta$ model parameter representing the average daily number of transmissions made by each infectious agent was estimated using the basic reproductive number estimates associated with COVID-19. $R_0$ represents the expected number of exposures resulting from each infectious case in a population where all agents are susceptible. A variety of values for $R_0$ have been found for COVID-19 depending on the variant of interest, but generally vary from around 2.5 (ancestral strain) to as high as 7 (Delta strain) or 10 (Omicron strain)~\cite{khandia_emergence_2022,burki_omicron_2022}. For the experiments in this paper, we aim to simulate an $R_0$ around 5, representing a strain that is more transmissible than the ancestral strain but not as transmissible as Omicron or Delta. 

The \textit{effective reproductive number}, $R$, associated with an infectious disease scenario is the actual number of exposures resulting from an infectious agent. This will vary over time based on susceptibility of agents in the population and mitigation strategies employed. The effective reproductive number at the beginning of the epidemic when most agents are susceptible is equivalent to $R_0$. 

\begin{figure}
    \centering 
    \def\scale{1}
    \begin{subfigure}{0.7\textwidth}
        \centering 
        \includegraphics[width=\scale\textwidth]{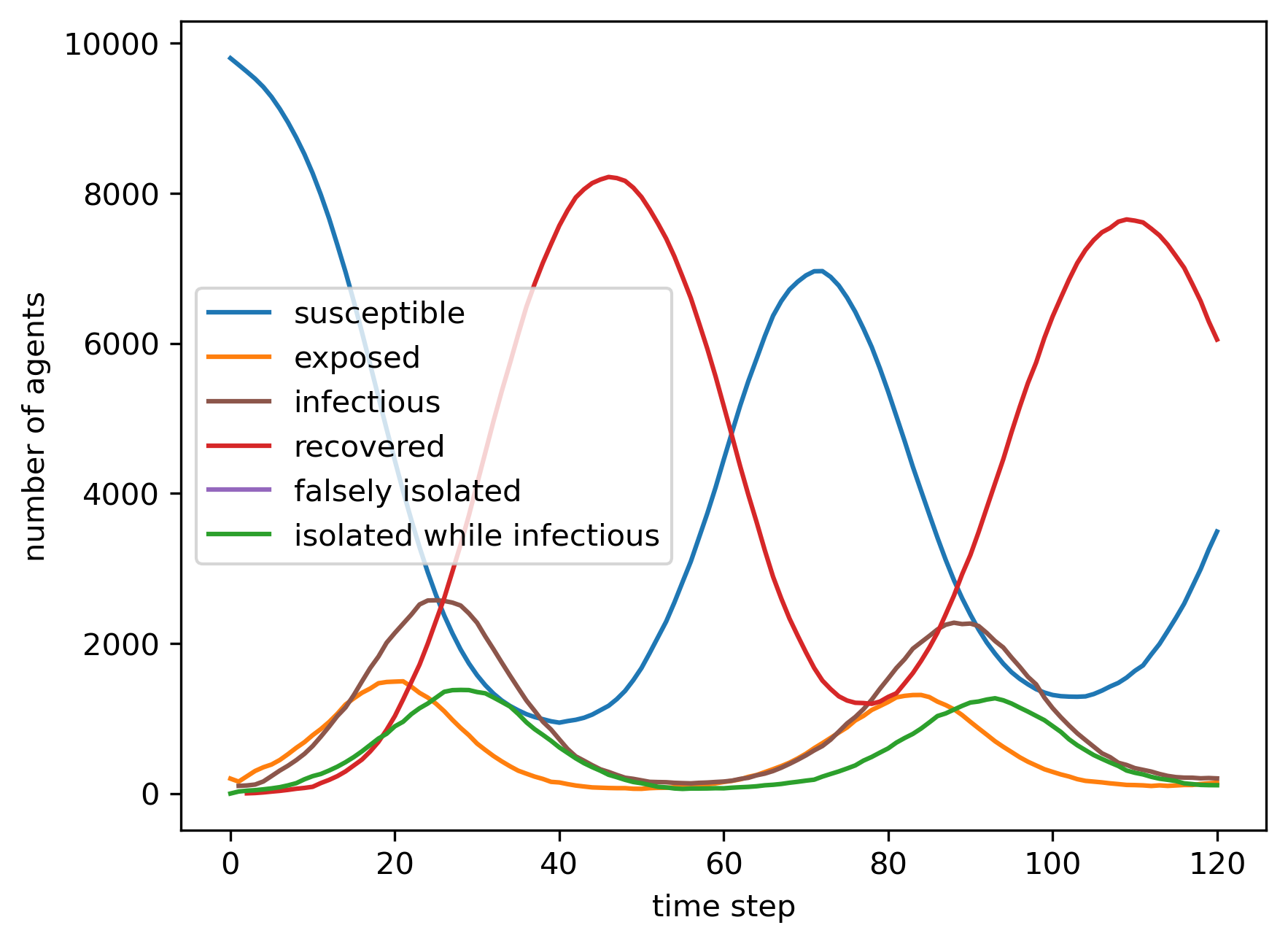}
        \subcaption{Disease dynamics}
    \end{subfigure} \\
    \begin{subfigure}{0.49\textwidth}
        \centering 
        \includegraphics[width=\scale\textwidth]{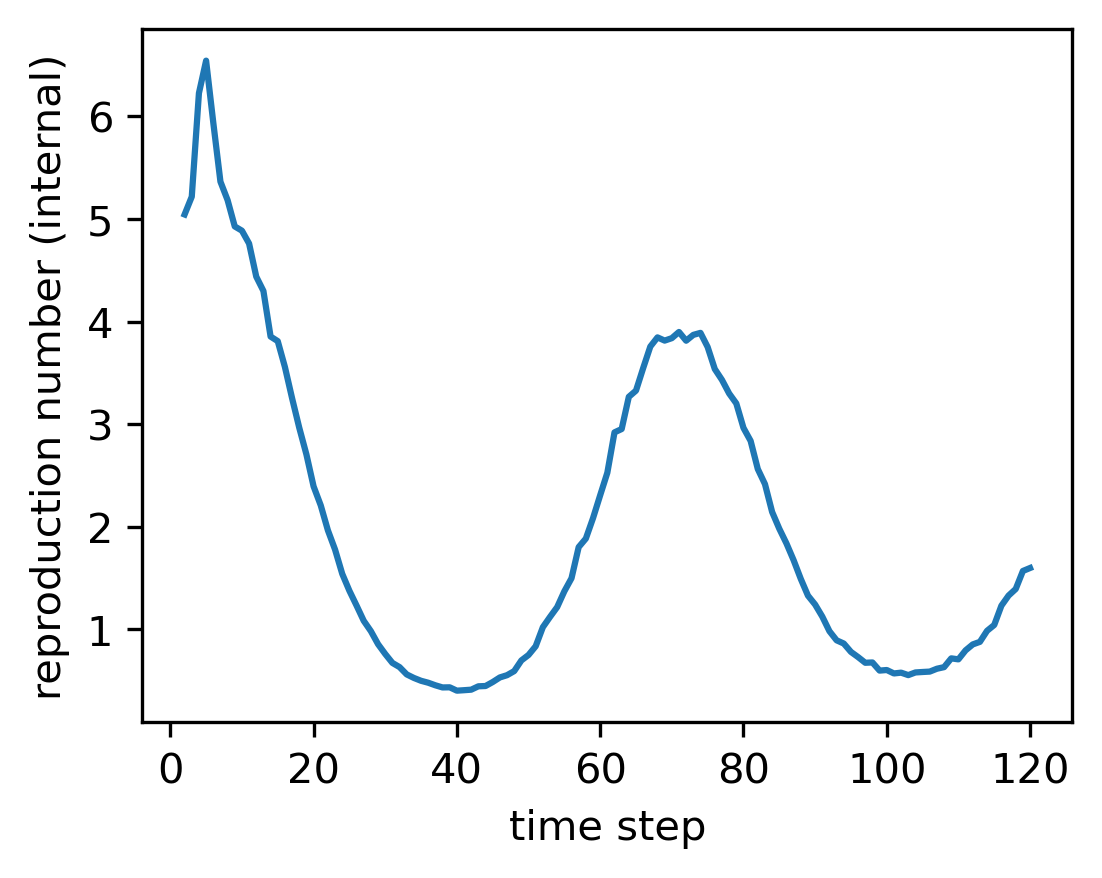}
        \subcaption{Effective reproductive number}
    \end{subfigure}
    \caption{\textbf{Baseline scenario with $\beta=0.4$.}}  
    \label{fig:baseline}
\end{figure}

To estimate the $\beta$ parameter, we first create a baseline scenario without intervention (no vaccination or testing). All other parameters (besides $\beta$) are fixed to their values in~\ref{app:params}. The estimated number of exposure-causing contacts associated with agents internal to the population is given by the second term of Eq.~\ref{eq:su-exp} (Eq.~\ref{eq:sv-exp} can be ignored in this case since $S_v=0$). The number of new internal exposures at time step $t$ is then given by
\begin{align}
    E'_i(t) = \beta \frac{I(t-1)}{P(t-1)} \cdot S_u(t-1).
\end{align}
It follows that the average number of exposure causing contacts at time step $t$ per infectious agent is equivalent to $\frac{E'_i(t)}{I(t-1)}$. The effective reproductive number at time step $t$ can be estimated by 
\begin{align}
    R_t = \frac{E'_i(t)}{I(t-1)} \cdot \tau_I,
    \label{eq:Rt}
\end{align}
where $\tau_I$ is the expected amount of time an agent is infectious. Following the model in Section~\ref{sub:vlm} $\tau_I$ can be estimated as, 
\begin{align}
    \tau_I &= \sigma_s \; \mathbb{E}\left[t_0 + t_P + t_S + t_F\right] + (1-\sigma_s) \; \mathbb{E}\left[t_0 + t_P + t_F\right] \\ 
    &= \mathbb{E}\left[t_0\right] + \mathbb{E}\left[t_P\right] + \mathbb{E}\left[t_F\right] + \sigma_s \; \mathbb{E}\left[t_S\right] \\ 
    &= 3 + (1.5 + 0.5) + 6.5 + (0.5)(1.5) = 12.25.
\end{align} 
This uses the fact that $\sigma_s=0.5$ and the mean of the Gamma distribution $\Gamma(k,\theta) = k\theta$. Plugging this value into Eq.~\ref{eq:Rt} and simplifying factors gives 
\begin{align}
    R_t = \frac{\beta S_u(t-1)}{P(t-1)} \cdot 12.25.
\end{align}
Assuming $S_u(t-1)/P(t-1) \approx 1$ and solving for $R_t=5$ gives a $\beta$ value around 0.4. The disease dynamics and effective $R$ value for the baseline scenario with $\beta=0.4$ are given in Fig.~\ref{fig:baseline}. 

\section{Simulation results: Supplemental figures for vaccination scenario A} \label{app:resultsA}

\begin{figure}[H] 
    \centering 
    \def\scale{1.0}
    \def\subscale{0.49}
    \begin{subfigure}[t]{\subscale\textwidth}
        \centering 
        \includegraphics[width=\scale\textwidth]{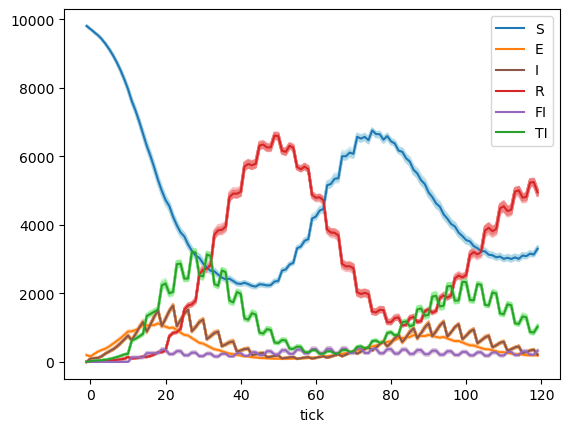}
        \subcaption{4 day testing interval, no pooling.}
    \end{subfigure}
    \hfill
    \begin{subfigure}[t]{\subscale\textwidth}
        \centering 
        \includegraphics[width=\scale\textwidth]{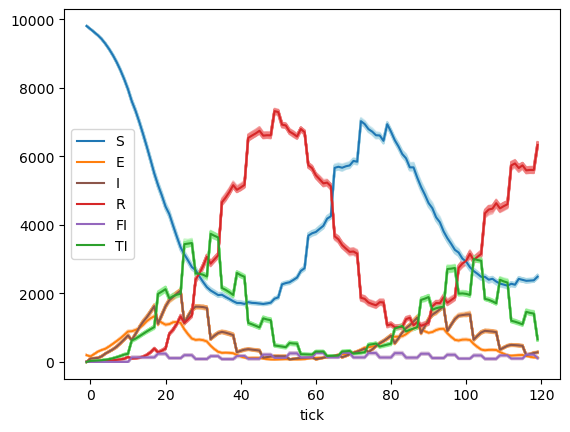}
        \subcaption{7 day testing interval, no pooling.}
    \end{subfigure}
    \\
    \begin{subfigure}[t]{\subscale\textwidth}
        \centering 
        \includegraphics[width=\scale\textwidth]{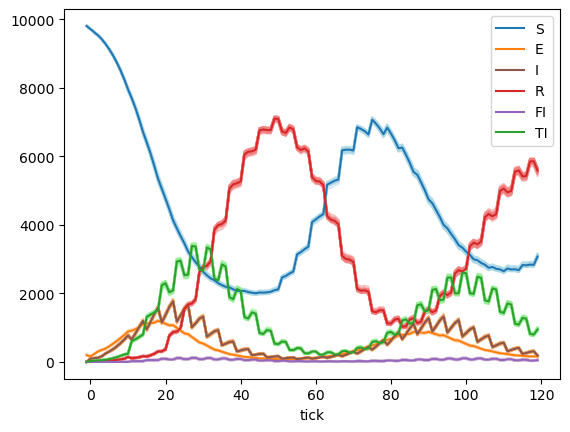}
        \subcaption{4 day testing interval, pool size 5.}
    \end{subfigure} 
    \hfill 
    \begin{subfigure}[t]{\subscale\textwidth}
        \centering 
        \includegraphics[width=\scale\textwidth]{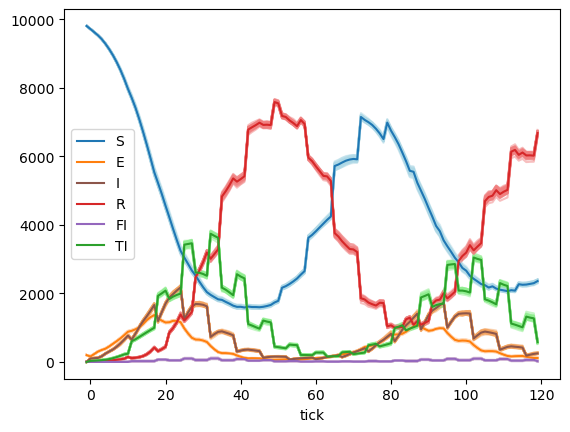}
        \subcaption{7 day testing interval, pool size 5.}
    \end{subfigure} 
    \caption{\textbf{Vaccination A (no vaccination) + Test A.} Subfigures show the number of individuals in the population occupying each of the disease model compartments (susceptible, exposed, infectious, recovered, falsely isolated, and isolated while transmitting). Plots for all 50 runs of each scenario are overlayed on the corresponding subfigure with the average of each compartment over all runs shown in a darker shade.}
    \label{fig:testA-vaxA}
\end{figure}

\begin{figure}[H]
    \centering 
    \def\scale{1.0}
    \def\subscale{0.49}
    \begin{subfigure}[t]{\subscale\textwidth}
        \centering 
        \includegraphics[width=\scale\textwidth]{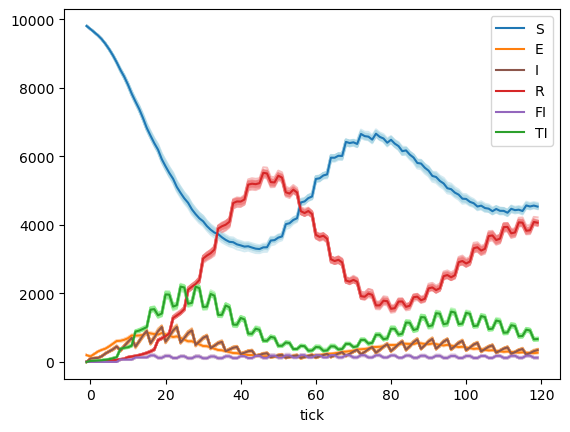}
        \subcaption{4 day testing interval, no pooling.}
    \end{subfigure}
    \hfill
    \begin{subfigure}[t]{\subscale\textwidth}
        \centering 
        \includegraphics[width=\scale\textwidth]{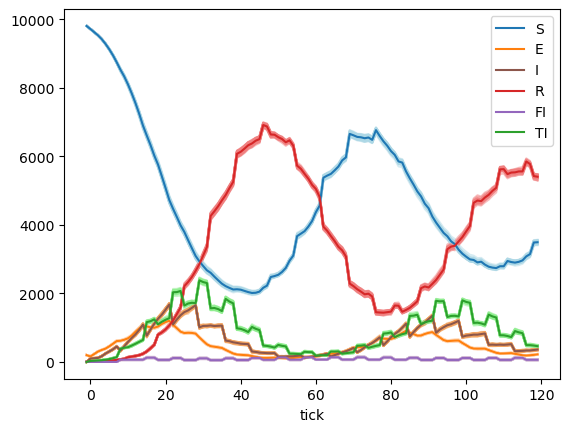}
        \subcaption{7 day testing interval, no pooling.}
    \end{subfigure}
    \\
    \begin{subfigure}[t]{\subscale\textwidth}
        \centering 
        \includegraphics[width=\scale\textwidth]{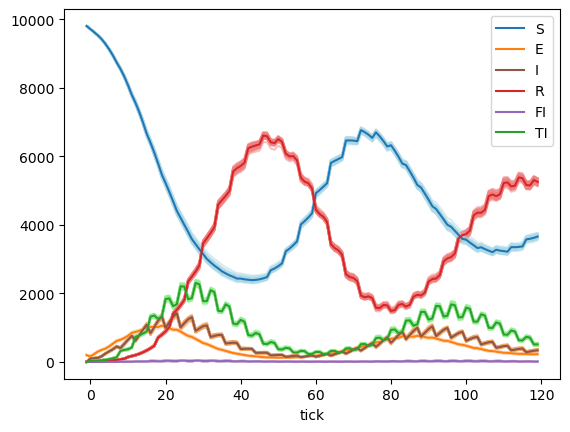}
        \subcaption{4 day testing interval, pool size 5.}
    \end{subfigure} 
    \hfill 
    \begin{subfigure}[t]{\subscale\textwidth}
        \centering 
        \includegraphics[width=\scale\textwidth]{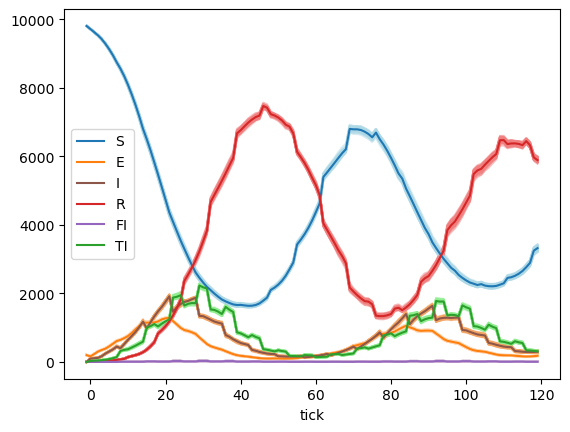}
        \subcaption{7 day testing interval, pool size 5.}
    \end{subfigure} 
    \caption{\textbf{Vaccination A (no vaccination) + Test B.} Subfigures show the number of individuals in the population occupying each of the disease model compartments (susceptible, exposed, infectious, recovered, falsely isolated, and isolated while transmitting). Plots for all 50 runs of each scenario are overlayed on the corresponding subfigure with the average of each compartment over all runs shown in a darker shade.}
    \label{fig:testB-vaxA}
\end{figure}

\section{Simulation results: Vaccination scenarios B and C} \label{app:resultsBC}
\begin{figure}[H]
    \centering
    \includegraphics[width=\textwidth]{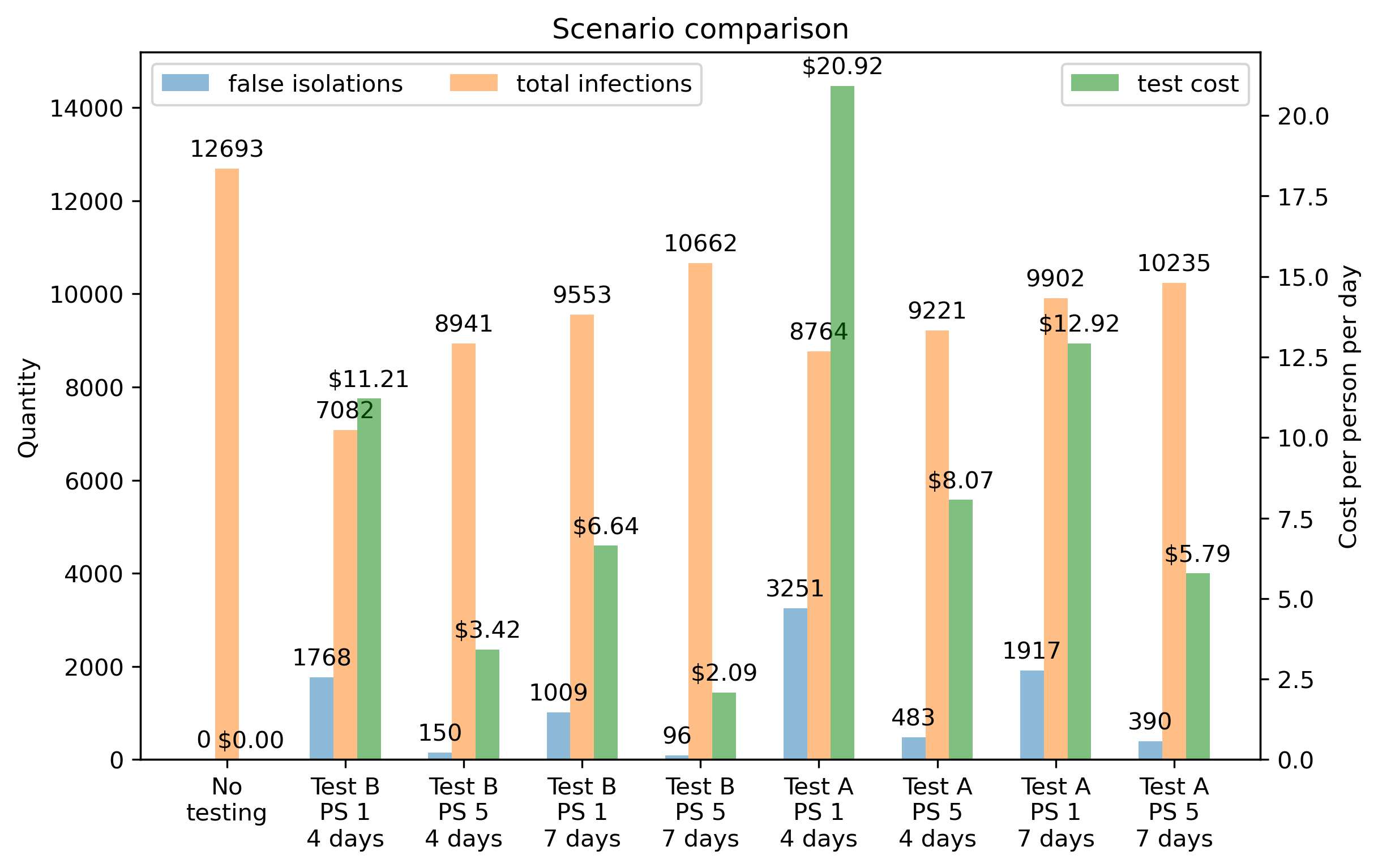}
    \caption{\textbf{Comparison of testing scenarios in a population with initial vaccine rollout (Vaccination B)}. Left and center bars (number of false isolations and total infections) correspond to the y-axis on the left, while the right-most bars (test cost) correspond to the y-axis on the right.}
    \label{fig:vaxB}
\end{figure}

\begin{figure}[H]
    \centering
    \includegraphics[width=\textwidth]{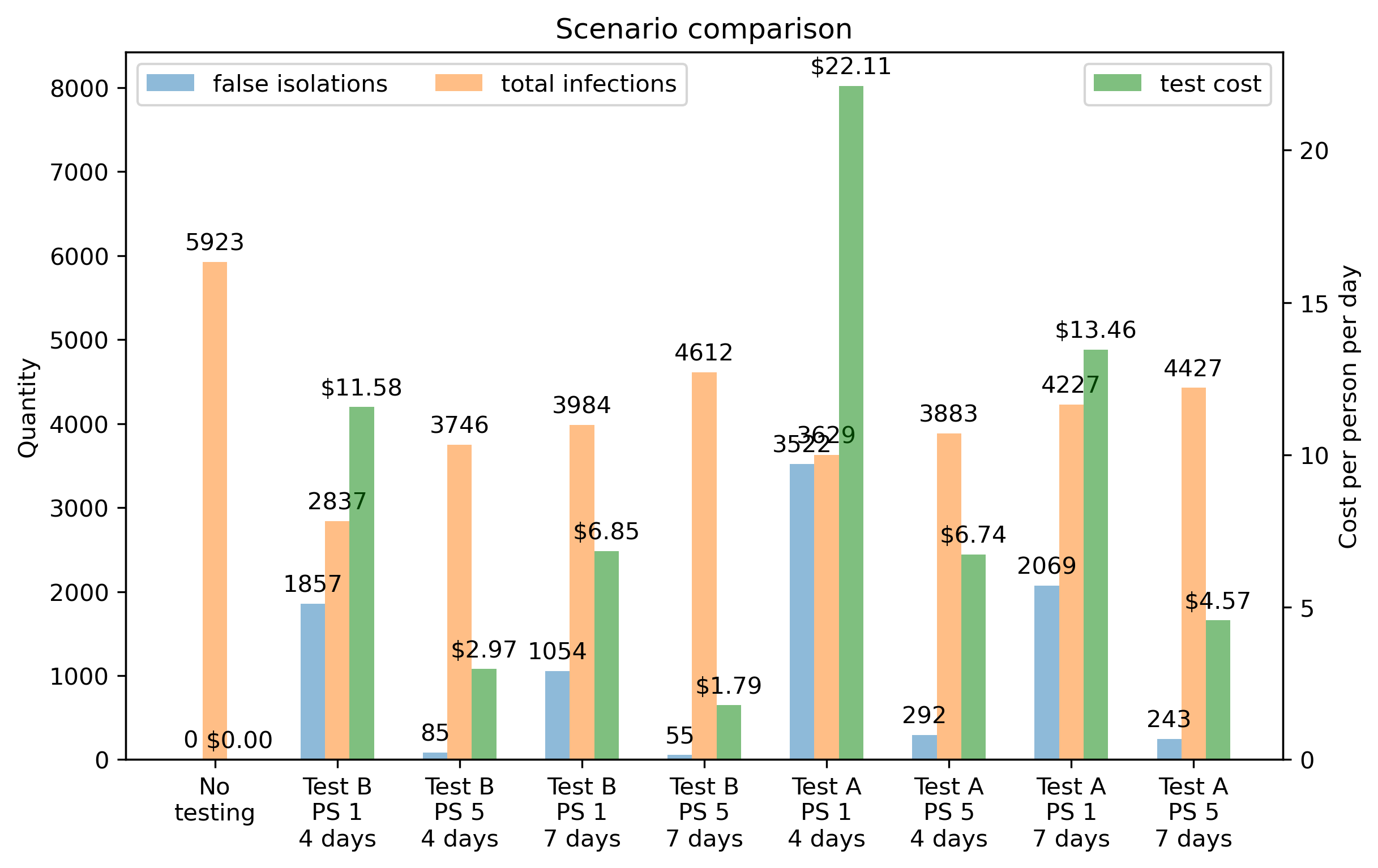}
    \caption{\textbf{Comparison of testing scenarios in a partially immunized population with continued vaccination rollout (Vaccination C)}. Left and center bars (number of false isolations and total infections) correspond to the y-axis on the left, while the right-most bars (test cost) correspond to the y-axis on the right.}
    \label{fig:vaxC}
\end{figure}

For completeness, results of simulations associated with vaccination scenarios B and C (Table~\ref{table:vax}) are shown here. Figs.~\ref{fig:vaxB} and~\ref{fig:vaxC} provide comparisons across all scenarios associated with vaccination status B and C, respectively. Scenario labels correspond to the type of test used, pool size used for pooled testing, and testing interval, respectively. Testing scenario cost is based on a cost of \$100 for test A and \$50 test B. Total testing costs are divided by 120 days and 10,000 people in the population to get the cost per person per day.
\vspace{1in}

\section{Summary Comparision with the Generalized SEIR Model} \label{app:gSEIR}
In 2020, Liangrong Peng \textit{et al.} introduced their Generalized SEIR model, 
which extended the classical SEIR model through the addition of 3 new states 
\cite{peng2020}. In particular, the Generalized SEIR model partitions a 
population of size $N$ into the following seven states:
\begin{itemize}[leftmargin=0.55in, noitemsep]
    \item[$S(t)$:] Susceptible cases
    \item[$P(t)$:] Insusceptible (\textit{i.e.} immune or vaccinated) cases 
    \item[$E(t)$:] Exposed, but not yet infectious, cases
    \item[$I(t)$:] Infectious cases
    \item[$Q(t)$:] Quarantined cases
    \item[$R(t)$:] Recovered cases
    \item[$D(t)$:] Closed (\textit{i.e.} deceased) cases 
\end{itemize}
E. Cheynet added the Generalized SEIR model to the MATLAB code base in 2020
\cite{cheynet2020}, and Felin Wilta \textit{et al.} utilized Cheynet's MATLAB 
implementation of the Generalized SEIR model in their 2022 paper on the 
death and recovery rates of COVID-19 \cite{wilta2022}. \modelabrev{} extends
the functionality of the Generalized SEIR model in a number of noteworthy 
ways, including:
\begin{itemize}
    \item The Generalized SEIR model is a dynamical (\textit{i.e.} ODE) system
        which models overall trends for a large population. As an ABM,
        \modelabrev{} models individual agents' interactions giving it the 
        flexibility to model a large variety of populations sizes and 
        characteristics.
    \item \modelabrev{} is able to separately model the impact of asymptomatic 
        versus symptomatic infectious agents on the spread of an infectious 
        disease.
    \item In the Generalized SEIR model, Insusceptible cases ($P(t)$) stay 
        insusceptible indefinitely. Whereas instead of having a separate 
        Insusceptible compartment, \modelabrev{} has two Susceptible states, 
        vaccinated and unvaccinated. Individual susceptible agents in the 
        vaccinated compartment still
        have a (potentially) non-zero probability of becoming infectious when 
        exposed to the disease. The probability of a vaccinated susceptible 
        agent can be tuned to match a given scenario.
    \item Similarly, there is no term in the differential equation for $R(t)$
        in the Generalized SEIR model to account for loss of immunity by 
        recovered cases. Conversely, \modelabrev{} is able to model loss of 
        immunity by individual recovered agents. 
    \item \modelabrev{} can model a variety of disease testing scenarios 
        (including pooled testing), and can model the impact of false negative
        and false positive tests.
    \item Under the Generalized SEIR model, only infectious cases can become 
        quarantined cases, and only quarantined cases can become recovered or
        deceased. In comparison, infectious agents under \modelabrev{} can 
        become recovered without isolating. Infectious agents who test positive 
        for the disease and healthy agents who receive a false positive for the 
        disease can be moved to isolation. Infectious agents in isolation are 
        moved to the recovered bin after their isolation period is complete, and 
        healthy agents in isolation are moved to one of the two susceptible bins 
        at the end of their isolation period.
    \item \modelabrev{} allows a user to specify a stochastic viral load profile
        for a given disease.
\end{itemize}

\end{document}